\documentclass{article}
\usepackage{spconf,amsmath,graphicx,hyperref}
\usepackage{tabularray}
\usepackage{rotating}
\usepackage{placeins}

\title{Music Source Restoration with Ensemble Separation\\ and Targeted Reconstruction}
\name{Xinlong Deng , Yu Xia , Jie Jiang}
\address{College of Artificial Intelligence, China University of Petroleum, Beijing\\
\{2022012257, 2023011703\}@student.cup.edu.cn, jiangjie@cup.edu.cn}
\begin{document}
%\ninept
\maketitle

\makeatletter
\renewcommand{\thefigure}{\arabic{figure}}
\renewcommand{\theHfigure}{\arabic{figure}} % for hyperref anchors
\@removefromreset{figure}{section}           % stop resetting per section
\makeatother

\begin{abstract}
The Inaugural Music Source Restoration (MSR) Challenge targets the recovery of original, unprocessed stems from fully mixed and mastered music. Unlike conventional music source separation, MSR requires  reversing complex production processes such as equalization, compression, reverberation, and other real-world degradations. To address MSR, we propose a two-stage system. First, an ensemble of pre-trained separation models produces preliminary source estimates. Then a set of pre-trained BSRNN-based restoration models performs targeted reconstruction to refine these estimates. On the official MSR benchmark, our system surpasses the baselines on all metrics, ranking second among all submissions. The code is available at https://github.com/xinghour/Music-source-restoration-CUPAudioGroup
\end{abstract}
\begin{keywords}
Music Source Restoration, Music Source Separation, Ensemble, Generation
\end{keywords}
\section{Introduction}
\label{sec:intro}

The Music Source Restoration (MSR) task, introduced in the ICASSP 2026 Signal Processing Grand Challenge, aims to recover original, unprocessed instrument stems from fully mixed and mastered music by effectively inverting production effects such as equalization, dynamic range compression, reverberation, and codec artifacts. While previous works target separation or restoration independently, MSR requires a unified approach. Music source separation (MSS) methods like BS-RoFormer \cite{lu2024music} assume linearly mixed clean sources, whereas restoration models like Apollo \cite{li2025apollo} and SonicMaster \cite{melechovsky2025sonicmaster} suppress degradations without explicitly recovering individual stems. MSR is more challenging than either task, as it demands the joint recovery of individual instruments from mixtures subjected to diverse, unknown, and often non-linear production chains.
% Previous works target only for separation or restoration. SCNet\cite{tong2024scnet} and BS-Roformer\cite{lu2024music}, as music source separation (MSS) methods, focus on demixing linearly mixed clean sources. Apollo\cite{li2025apollo} and SonicMaster\cite{melechovsky2025sonicmaster}, as audio restoration methods, operate to suppress degradations without explicitly recovering individual stems. In contrast, MSR jointly requires restoring and separating multiple overlapping instruments under diverse and often unknown production chains, making it substantially more ill-posed and challenging than either MSS or conventional restoration alone.

In this paper, we propose a two-stage system for MSR. Our approach first leverages an ensemble of pre-trained MSS models to generate initial source estimates for each instrument. These estimates are subsequently processed by a BSRNN-based restoration module \cite{zang2025msrbench} for joint separation refinement and targeted restoration. Experimental results on the official MSR benchmark demonstrate that our system outperforms the official baselines across all three objective tracks, ranking second on the ICASSP 2026 MSR leaderboard.

% In this paper, we propose a two-stage system for MSR. Our approach first applies an ensemble of pre-trained MSS models to generate preliminary separation results for each instrument, followed by a BSRNN-based restoration module trained to map degraded mixtures to unprocessed stem targets, performing generative-like re-separation and targeted restoration. Experimental results on the official MSR benchmark show that the proposed system improves upon the official baselines across the three objective tracks and achieves the 2nd overall rank in the leaderboard.

\begin{table*}[t]
\centering
\caption{Results of different MSR systems per instrument class.}
\label{tab:result}
\begin{tblr}{
  cells = {c},
  cell{2}{1} = {r=2}{},
  cell{4}{1} = {r=2}{},
  cell{6}{1} = {r=2}{},
  hline{1-2,4,6,8} = {-}{},
}
Model             & Metric & Vocals          & Gtr.            & Key.            & Synth           & Bass            & Drums           & Perc.           & Orch.           \\
BSRNN (baseline)   & MMSNR  & \textbf{1.3365}          & 0.2722          & 0.0588          & 0.0223          & 0.6303          & 0.8569          & 0.0000               & 0.0388          \\
                  & FAD    & 0.3476          & 0.4085          & 1.0690           & 0.9027          & 0.7334          & 0.6393          & 1.1880           & 0.7472          \\
EnsembleSep       & MMSNR  & 1.3047          & 0.5836          & 0.2578          & \textbf{0.0693} & \textbf{1.4700}   & 1.6712          & \textbf{0.6846} & 0.1348          \\
                  & FAD    & \textbf{0.2607} & 0.4832          & 0.7842          & 0.8156          & 0.4387          & 0.4540           & \textbf{0.7412} & 0.7460           \\
EnsembleSep+BSRNN & MMSNR  & 1.3298 & \textbf{0.6274} & \textbf{0.4077} & 0.0596          & 1.3800            & \textbf{1.9461} & 0.0000               & \textbf{0.1570}  \\
                  & FAD    & 0.2680           & \textbf{0.2802} & \textbf{0.6662} & \textbf{0.7525} & \textbf{0.4045} & \textbf{0.3236} & 0.8461          & \textbf{0.5981} 
\end{tblr}
\end{table*}

\section{METHODOLOGY}
\label{sec:format}

\subsection{Ensembled Music Source Separation}
\label{ssec:subhead1}

\begin{figure}[h]
\centering
\begin{minipage}[t]{0.48\textwidth}
  \vspace{0pt}
  \includegraphics[width=\linewidth]{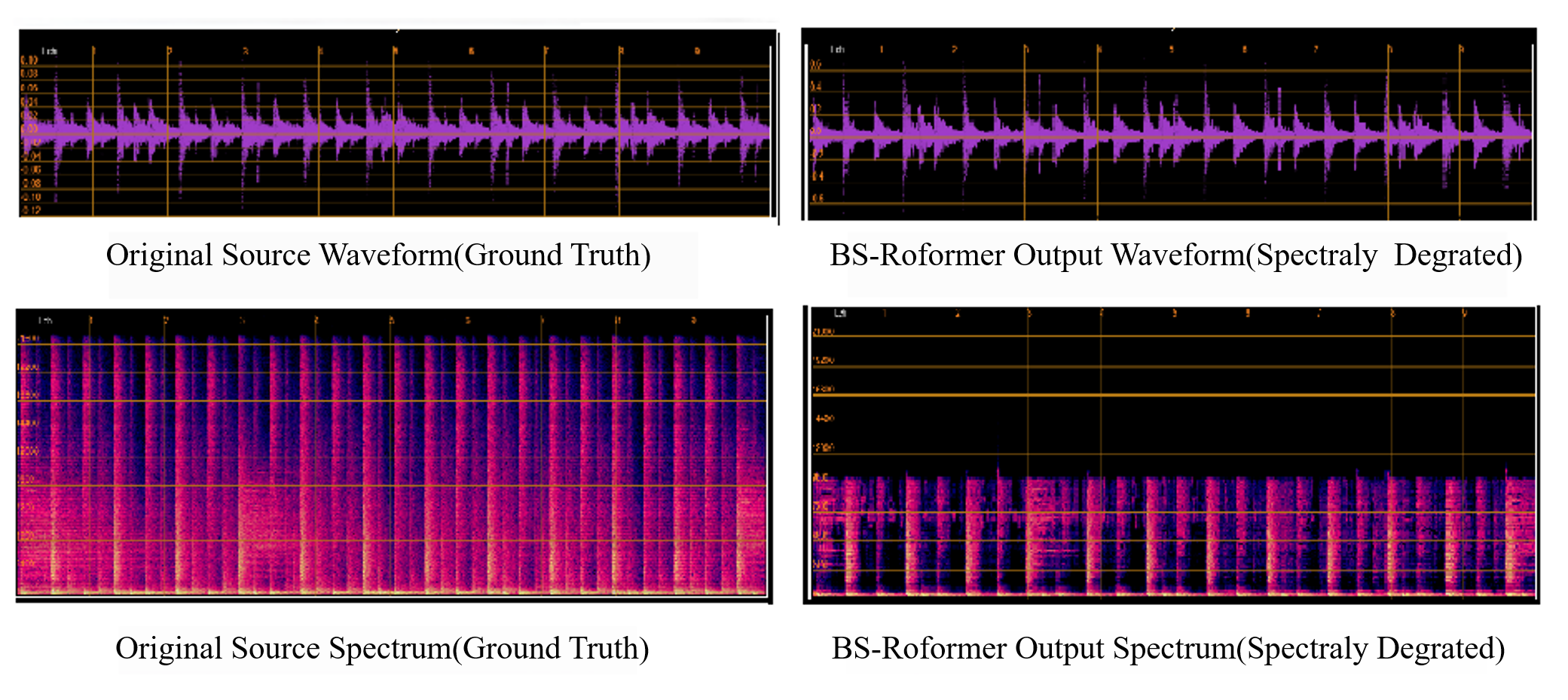}
\end{minipage}
\caption{Waveform and spectrogram of the original audio and the BS-RoFormer separated output for drums.}
\label{fig:bs-roformer-results}
\end{figure}

Although MSS models are not explicitly designed to recover clean sources from degraded mixtures, our experiments indicate that training with mask target enables them to learn instrument-specific spectral and structural priors. As shown in Figure \ref{fig:bs-roformer-results}, the separators trained on clean mixtures retain the ability to localize and extract residual instrument components from degraded mixtures, provided their discriminative structures remain partially intact despite the degradation.

Motivated by this observation, we implement a cascaded ensemble to cover eight instrument classes. We first utilize BS-RoFormer \cite{lu2024music} to generate initial estimates for vocals, bass, drums and an aggregate other stem. Given that percussion is often entangled with drums, we then employ MDX23C \cite{mezza2024benchmarking}, a specialized drum separation model, to refine the drum track and isolate percussion. Any remaining instrumental content is then categorized under the other stem. 

However, as MSS models are generally trained on clean mixtures, they face a domain gap when applied to mastered music. Consequently, their outputs often fail to account for complex production effects, resulting in incomplete source disentanglement and persistent audio degradations. This necessitates a dedicated second stage designed specifically for joint separation refinement and restoration.

% However, since both BS-Roformer and MDX23C are trained on clean mixures, the separation results exhibit incomplete source disentanglement and the extracted sources remain degraded, requiring refined separation and restoration stages.

\subsection{Targeted Source Restoration}
\label{ssec:subhead2}

Conventional audio restoration models typically assume that the input signal is already isolated \cite{li2025apollo, melechovsky2025sonicmaster}. Consequently, these models cannot be directly applied to the outputs of our system's first stage due to a fundamental domain mismatch between the degradation patterns in isolated sources and the artifacts present in the separation estimates. To bridge this gap, we require a framework capable of treating these imperfect estimates as informative priors. This allows for joint separation refinement and restoration, aimed at recovering the original, unprocessed instrument stems.

In the absence of existing specialized models for this task, we adopt the baseline system provided by the MSR Challenge: a suite of pre-trained BSRNN-based MSR models \cite{zang2025msrbench}. These models are specifically designed and trained to map degraded mixtures to unprocessed stem targets by performing simultaneous separation and restoration. In our pipeline, we utilize these models for targeted refinement, i.e., the separation estimates from the first stage are fed into their respective instrument-specific BSRNN models. This allows the system to leverage the initial estimates as informed starting points to further suppress residual artifacts and achieve better restoration of each instrument stem.

\section{RESULTS}
\label{sec:pagestyle}
Table~\ref{tab:result} summarizes the performance of the baseline system, the ensembled MSS (EnsembleSep), and the proposed two-stage MSR system (EnsembleSep+BSRNN) on the MSRBench validation set. Evaluation is conducted using Multi-Mel Spectrogram Signal-to-Noise Ratio (MMSNR) and Fréchet Audio Distance with CLAP (FAD-CLAP) \cite{kilgour2018fr} to assess time-frequency accuracy and semantic restoration respectively. In general, EnsembleSep+BSRNN  achieves superior performance across most instrument classes. However, for Percussion, the MMSNR for both the baseline and our system approached zero. This is due to a significant distribution shift and data scarcity between the training and validation sets. Consequently, the BSRNN stage was bypassed for this specific instrument class during evaluation on the test set.

Our proposed system further demonstrated its efficacy by outperforming the baseline on the Challenge test set. Specifically, it achieved scores of 2.3405 (MMSNR), 0.2253 (FAD), and 0.0164 (Zimt) \cite{alakuijala2025}, alongside an average MOS of 3.2262.

\section{CONCLUSION}
\label{sec:typestyle}

We propose a two-stage system for MSR that integrates ensemble separation with targeted reconstruction. Experimental results validate the efficacy of this system. Furthermore, our analysis reveals that data scarcity remains a significant bottleneck; we posit that scaling with higher-quality, diverse training data will be essential to further advancing the performance of separation and restoration modules.

% Below is an example of how to insert images. Delete the ``\vspace'' line,
% uncomment the preceding line ``\centerline...'' and replace ``imageX.ps''
% with a suitable PostScript file name.
% -------------------------------------------------------------------------
% \begin{figure}[htb]

% \begin{minipage}[b]{1.0\linewidth}
%  \centering
%  \centerline{\includegraphics[width=8.5cm]{image1}}
%  \vspace{2.0cm}
%  \centerline{(a) Result 1}\medskip
% \end{minipage}
%
% \begin{minipage}[b]{.48\linewidth}
%  \centering
%  \centerline{\includegraphics[width=4.0cm]{image3}}
%  \vspace{1.5cm}
%  \centerline{(b) Results 3}\medskip
% \end{minipage}
% \hfill
% \begin{minipage}[b]{0.48\linewidth}
%  \centering
%  \centerline{\includegraphics[width=4.0cm]{image4}}
%  \vspace{1.5cm}
%  \centerline{(c) Result 4}\medskip
% \end{minipage}
%
% \caption{Example of placing a figure with experimental results.}
% \label{fig:res}
%
% \end{figure}

%\vfill
%\pagebreak


\makeatletter
\renewcommand{\refname}{REFERENCES}
\patchcmd{\thebibliography}{\section*{\refname}}{\section{\refname}}{}{}
\makeatother

{\footnotesize
\bibliographystyle{IEEEbib}
\bibliography{strings,refs}
}

\end{document}